\documentstyle [11pt,aaspp4,psfig]{article}

\begin{document}
\newcommand{\figureout}[3]{\psfig{figure=#1,width=7in,angle=#2} 
    \figcaption{#3} }
\title{RXTE observation of PSR B1706-44 and implications for
theoretical models of pulsar emission}

\author{A. Ray \altaffilmark{1}, A.K. Harding}
\affil{NASA/Goddard Space Flight Center, Greenbelt, MD 20771} 
\author{M. Strickman}
\affil{Naval Research Laboratory, Space Sciences Division, Washington,
D.C. 20375}

\altaffiltext{1}{On leave of absence from: 
Tata Institute of Fundamental Research, Mumbai 400005, India; E-mail:
akr@tifrvax.tifr.res.in}



\begin{abstract}

We report on results of an observation with the Rossi X-Ray Timing
Explorer (RXTE)
of PSR B1706-44 with a live time of 132 ks, 
to search for pulsed X-ray emission.  PSR B1706-44 is a
radio
and high-energy gamma-ray pulsar (detected by EGRET), but no pulsed
emission
has been detected in the X-ray band.  Since most of the other known
gamma-ray
pulsars emit pulsed X-rays, it is expected that PSR B1706-44 would also be
an
X-ray pulsar.  However, while the ROSAT PSPC detected a source at the
pulsar position, it did not detect pulsations, giving a pulsed fraction
upper
limit of 18\%.  The RXTE observations to search for modulation at
the pulsar period 
were carried out in November 1996 and May 1997
during the low states of the nearby X-ray binary 4U1705-44.
No significant modulation was detected at the pulsar period, giving
an upper limit of $10^{-6} \rm \; photons \; cm^{-2} \; s^{-1} \; keV^{-1}$ in
the interval 9 keV $\leq E \leq 18.5$ keV.
The implications of this upper limit of the pulsed flux from the RXTE
observation, taken together with multiband observations of this pulsar
are examined in the context of theoretical models
of pulsar particle acceleration zones and associated 
high energy electromagnetic emission.
\end{abstract}


\section{Introduction} 

In recent years multiwavelength observations of several pulsars
are now allowing the
testing and refinement of theoretical models of 
pulsar emission mechanisms and particle acceleration responsible for
high energy electromagnetic radiation.
There are now seven pulsars (eight if PSR0656+14 is counted)
which are to emit {\it pulsed} gamma-rays observed by the EGRET 
instrument aboard
the Compton Gamma-ray Observatory (CGRO) (see Thompson et al 1997),
. 
In addition there are more than two dozen isolated (rotation powered)
neutron stars which are detected by a number of X-ray satellites
such as Einstein, ROSAT and ASCA (see e.g. Becker and Tr\"umper 1997).
Spectra, pulse shapes, relative arrival times of pulses in different
electromagnetic bands and, in some cases, phase resolved spectra
offer information which can be useful for identifying
the zone of acceleration
of the charged particles in the magnetosphere surrounding the pulsar,
and the way these particles produce and interact with electromagnetic
radiation. However, of the seven gamma-ray pulsars identified, only
six have detected pulsations in the X-ray band, and only five (Crab, Vela,
Geminga, PSR1055-52 and PSR1951+32) have measured X-ray spectra. 
No X-ray pulsations have been
detected from the gamma-ray pulsar PSR B1706-44, although steady X-ray
emission has been seen by ROSAT from this object. Additionally, ASCA
has taken spectral data of the nebular emission surrounding the pulsar
and a combined analysis of this and the higher spatial
resolution (single-band)
ROSAT HRI data has been performed by Finley et al (1998). 

PSR B1706-44
is a young neutron star (spin down age 1.75$\times 10^4$ yr) with a 
spin period of 0.102 s and has a large spin down luminosity of
3.4$\times 10^{36} \rm erg \; s^{-1}$ at a distance of 1.82 kpc.
The pulsar which 
was originally discovered in the
radio band by Johnston et al (1992),
ranks fourth in a rank ordered list of spin-down flux 
($ I \Omega \dot\Omega/D^2$).
The soft X-ray emission was
detected by the ROSAT PSPC (Becker et al 1992) with an upper limit to
modulation of the X-ray flux at the 18\% level (Becker, Brazier and Trumper
1995). As most gamma-ray pulsars emit pulsed X-rays, it is of interest
to detect the X-ray pulsations from this high spin-down flux pulsar
and to study its pulsed spectrum. In this paper we report on our
attempt to detect X-ray pulsations from this pulsar using the
Rossi X-ray Timing Explorer (RXTE). Preliminary
results from this observation were reported by Ray, Harding and
Strickman (1997).
Although we have not detected any X-ray pulsations, the stringent
upper limit to the pulsed flux we obtained has implications for
magnetospheric particle acceleration and radiation models. In section 2,
we describe the characteristics of PSR B1706-44 and the theoretical
motivations for this study. In section 3, we describe the observation
carried out with RXTE and its analysis. In section 4, we compare the 
broadband observations of this pulsar 
with the theoretical models of the pulsed radiation mechanism
in the light of our new observation. 
In section 5, we discuss our conclusions.

\section{Motivation for pulse detection and characteristics of PSR 1706-44}

X-ray emission from rotation powered pulsars can be a combination of
varying amounts of three spectral components: 1) a soft blackbody
2) a power law and 3) a hard component connected with heated polar caps.
Blackbody-like emission seen at
energies between 0.1 and 0.5 keV can result from surface cooling
of the neutron star.   Polar cap heating by energetic particles or internal
sources may produce a second thermal component at higher energies..
A power-law component could result from non-thermal radiation of
particles accelerated in the pulsar magnetosphere.

Measurement of the pulsed X-ray flux from polar cap heating
can act as a calorimeter of the returned particle flux from
a polar cap (PC) or outer gap (OG) vacuum discharge accelerator.
The returned particle flux is
an indicator of the properties and mechanisms of the non-thermal
(gamma-ray and power law X-ray) emission
from the pulsar. 
In particular PC models predict a
returned flux that is a  small percentage of the total accelerated
particle flux
(Harding and Muslimov 1998 (HM98), Arons 1981) 
whereas the OG models predict
roughly equal numbers of particles accelerated towards the PC
as flowing away 
(Halpern and Ruderman 1993; Wang et al 1998 (hereafter WRHZ)).
The latter, therefore, would give rise to 
comparable X-ray
and gamma-ray  photon fluxes. These measurements for pulsars could
serve as an important separator of the two classes of models.
 
At present, it is not clear how and where in the pulsar magnetosphere
the pulsed non-thermal high energy emission originates.  Polar cap models
assume that particles are accelerated above the neutron star surface
and that radiation results from a curvature radiation 
(Daugherty and Harding 1996)
or inverse-Compton 
(Sturner and Dermer 1994)
induced pair cascade in
a strong magnetic field.  Outer-gap models 
(Cheng, Ho and Ruderman, 1986; CHRa,b)
assume that acceleration occurs along null charge surfaces in the outer
magnetosphere and that the radiation results from curvature 
radiation/photon-photon pair production cascades
(Ray and Benford 1981).
The computed spectra of the pulsars in both polar cap 
(DH96) and outer-gap 
(Ho 1989) models predict that the spectrum will extend below
50 keV. In polar cap models this emission is primarily power-law
synchrotron emission that will terminate at the cyclotron energy of the local
magnetic field, which could be anywhere from $\sim$ 10 keV at the
neutron
star surface down to $\sim$ 0.1 keV several stellar radii above the
surface.
A measurement of the pulsed emission below 50 keV would determine how
far
the spectrum of the non-thermal component extends.  If a cutoff in the
spectrum is observed, it might be a measure of the local field strength
in the emission region and thus, a limit of the proximity of a polar cap
cascade to the neutron star surface.


The seven gamma-ray pulsars detected by the Compton Gamma-ray
Observatory
have high
spin down luminosities and, because of their association with
EGRET gamma-ray sources, probably have pulsed X-ray fluxes.
In the case of PSR B1706-44 there is not enough spectral
information to classify the  emission as thermal vs non thermal
nor determine whether any fraction is pulsed.
ROSAT has made pointed observations of
this pulsar on three different occasions with
both the PSPC (Feb 1992 and Sept 1992 -- see Becker et al 1995)
and the HRI (Mar 1995 -- see Finley et al 1998) 
for 11.8 ks and 28 ks respectively.
The former showed a net count rate of 0.022 $\pm$ 0.002 cps
for the energy range 0.1 - 2.4 keV.
The net count rate reported by HRI was 0.0087 cps
which was considered consistent with the PSPC count rate.
The HRI-determined J2000 source location was in good agreement
with the previous radio (interferometric and timing) positions
of Frail et al (1994) and Johnston et al (1992).
In addition, ASCA SIS and GIS instruments imaged the region
around PSR B1706-44 for net exposures of 20 and 23 ks respectively
on 10 Sept 1994 with reported net count rates of 0.016 and 0.018 cps
respectively.
While the 2$\sigma$ upper limit of the pulsed fraction is 18\%
in the ROSAT PSPC band, the upper limit for the ASCA band 
gives a 22\% limit in the 2.0-10.0 keV band at a 99\% confidence level.
The size of the synchrotron nebula (to the flux levels searched by
ROSAT HRI) was found to be $27^{\arcsec}$ in radius (Finley et al. 1998).

Spectral analysis of ROSAT PSPC data did not distinguish between 
black body and power law models, but the lack of pulsations and the
similarity of spin parameters of PSR B1706-44 with those
of Vela were used by Becker et al. (1995) to favor a power law model
with $dN /dE \propto E^{-2.4}$ which suggested  synchrotron emission
from a pulsar powered nebula as the likely origin of detected
(steady) X-rays. The corresponding neutral hydrogen column depth
$N_H$ was required to be 5.4 $\times 10^{21} \rm cm^{-2}$ from spectral
fits. The flux within the 0.1-2.4 keV range and the
bolometric luminosity (isotropic) at a distance of 1.8 kpc
are about  3.2 $\times 10^{-12} \rm \; erg \; s^{-1} \; cm^{-2}$
and 1.3 $\times 10^{33} \rm \; erg \; s^{-1}$ respectively. 
The energy spectra obtained with ASCA SIS and GIS are 
well fit by a single power law model which gives
a photon index $\alpha = 1.7 ^{ +0.5}_{-0.4}$ 
and $N_H = 1.9 \times 10^{21} \rm cm^{-2}$
with L(2-10 keV) = 5.3 $\times 10^{32} \rm \; erg \; s^{-1} \; cm^{-2}$.

The bright Low Mass X-ray Binary 4U 1705-44
is about $22^{\arcmin}$ away from PSR B1706-44 (see Fig 1).
In addition there is
a faint source in soft X-ray band at $RA = 17^h 10^m 44^s$
DEC=-44$^{o}$ 33$^{\arcmin}$ 28$^{\arcsec}$. These two sources 
cause confusion and background problems for
RXTE since they are all within the
FWHM response circle of the RXTE PCA detector. Additionally, 
the large scattering wings of the ASCA mirrors mean that flux
from these sources contributes to the ASCA background for PSR B1706-44.

The EGRET instrument shows a photon spectral index of 1.57 $\pm$ 0.05
in the gamma-ray energy range of 100 MeV to 2 GeV for this pulsar
and steepens at higher energies (Fierro 1995).
In addition there are upper limits to the optical (R-band) magnitude
of this pulsar which are discussed together with multiband observations
in the context of theoretical models in Section 4.3.

\section{RXTE Observation \& Analysis}

PSR B1706-44 was observed as an RXTE Target of Opportunity (TOO)
when the nearby bright LMXB 4U 1705-44,
went into a low state (corresponding to 5 ct/s in the All Sky Monitor).
This LMXB, which is about $22^{\arcmin}$ away from PSR B1706-44,
remains in the FOV of the PCA instrument and is therefore a source
of contamination which reduces the available pulsed signal to noise.
Hence it is advantageous to perform the experiment to detect pulsations
when the LMXB is in a low state. A gray-scale image of the region of the
sky downloaded from the High Energy Astrophysics Archive
(HEASARC) with a superimposed circle denoting the XTE PCA HWHM radius
is shown in Figure 1.

PSR B1706-44  was allocated 147 ksec of TOO observation. 
It was observed on two different occasions between
November 9-12, 1996 and May 16-19, 1997. 
The dates, instrument configurations,
exposures etc., are listed in Table 1.
After excising the times of
SAA passage and earth occultations etc., the on-source
live time was 132 ks. 
During this time all 5 Proportional Counter Units (PCU) were kept on.
On certain occasions, the nearby LMXB 4U1705-44 underwent type I X-ray
bursts, enhancing its luminosity by several orders of
magnitude for a few tens to a hundred seconds even though generally, the 
LMXB was in a quasiperiodic low state (as seen by the All Sky Monitor
as referred to above) for $\sim$ 20 days. Several hundred seconds
of data, comprising the bursts and a roughly  equal amount of time immediately
after each burst were excluded from our timing analysis for PSR B1706-44 X-ray
pulse detection by visually examining the Standard2 light curves
and then using the ftools software package {\it grosstimefilt}.

A preliminary investigation of the data revealed that the average
count rate in channels 5-22 
(in GoodXenon-16s mode)
was substantially higher than that in channels 23-49. 
It is also well known
that the instrumental background in channels 49 through 255 usually
is too high to make low signal to noise detections of pulsed emission
feasible. Hence we divided the data into bands containing
channels 5 through 22 and
23 through 49 and used mainly the latter set of channels for timing
analysis towards pulse detection because the
contaminating emission from the LMXB was lower in these channels. 
For the span of data which was of the mode EA-125us-64M-0-1s
in November 1996, we chose the appropriate channel boundaries (CPIX4)
which corresponded to the same range of energy as the chosen channels
(23-49) for the GoodXenon mode, for a combined analysis.


In order to ameliorate contamination from the LMXB, we considered
offset pointing from the pulsar away from the LMXB. However, we
calculated, and later confirmed with raster scans of the region,
that this would have resulted in a net decrease in the signal to noise
ratio.

The photon events from
the two separate Experiment Data Systems (PCA/EDS) 
configurations were separately analyzed
and fed into the ``fasebin'' ftool.
This package handles the solar system barycentric corrections of the
arrival times of the photons (via the JPL DE200 solar system and
the RXTE orbital ephemerides),
and then epoch-folds the data according to a supplied
pulsar ephemeris. The radio pulsar
ephemerides used in the epoch folding program, kindly supplied by
V. Kaspi et al, are displayed in Table 2.
The radio observations ruled out any sudden period jumps of
the pulsar spin period (glitches) near or in between the two
epochs of RXTE observations, which are separated by nearly six months.
Other ftools packages allow manipulation of the epoch folded phaseogram 
according to spectral channels, binning in adequate number of phase bins etc.
The epoch folded phaseogram of these events is displayed in
Figure 2.
The origin of the phaseogram corresponds to the absolute phase of
the radio pulse.
There is no compelling evidence for pulsed emission at the pulsar
period.  Note that the reduced $\chi^2$ per degree of freedom
(19 degrees of freedom) for the histogram displayed in Figure 2 is
1.99.
This means that a random set of fluctuations consistent with a
uniform background of the zero level shown (corresponding to 94.16 cps),
i.e. the absence of a pulsation at the radio folding period, is ruled
out
at a 99.25 \% confidence level.
Currently we can only obtain the following 2 $\sigma$
upper limit to the  {\it pulsed }
flux, based on the total observation of 132 ks,
in the energy interval 9 keV $<$ E$_{\nu}$ $<$ 18.5 keV band 
(Ulmer et al 1991):
\markcite{aray:ulmer91} :

$$ UL = { 2 \over \Delta E} {\sqrt{C_{tot}}\over (A_{eff} t)}
\sqrt{ {\beta \over 1 - \beta} }
 = 10^{-6} \rm \;photons\; cm^{-2}s^{-1}keV^{-1} $$

\noindent where we have taken the pulsar duty cycle $\beta$ = 1/2
and the RXTE effective area A$_{eff}$ to be 6000 cm$^2$.
This upper limit is shown together with the detections and
upper limits of PSR B1706-44 in other energy bands in 
Figure 3.

%

%
%
\section{Implications of multiband spectra for theoretical models:}

The photon spectrum shown in Figure 3 from the ROSAT (keV) band to the 
TeV band 
has implications for theoretical models of pulsar radiation
mechanisms. High energy electromagnetic radiation from pulsars
is due to relativistic electrons and positrons which are accelerated
in the magnetosphere surrounding the neutron star. 
The models for magnetospheric accelerators fall into two main classes 
depending upon their location with respect to the neutron star.  
In the polar cap model invoked originally to power radio emission from 
a pulsar (Ruderman and Sutherland 1975; Daugherty and Harding 1982) the 
acceleration zone for the primary electrons (positrons) is
relatively close to the neutron star surface. The outer magnetospheric
accelerators (Holloway 1973; -- the ``outer gap"), 
proposed for energetic photons (Cheng, Ho, \& Ruderman
1986a,b, hereafter CHRa and CHRb), accelerate particles
at a distance which is a good fraction of the light
cylinder radius from the neutron star. As extensive e$^{\pm}$ flows
are generated by both polar cap and outer-gaps, the pairs from one would be
expected to quench the other, and it is difficult to expect that both
could survive on the same field lines, unless the field lines threading
the two acceleration zones have differing curvatures and signs of net charge
density etc. (see Wang et al 1998 for a discussion). 

Currently, there are several different versions of the outer gap model
in the literature
which differ in significant detail as to the mechanisms of gap discharge
through e$^{\pm}$ pair production and the process of electromagnetic
radiation emitted by these relativistic particles. The classic version
of the CHR model is now extended by Zhang and Cheng (1997), Cheng and
Ding (1994) as well as by Wang et al (1998). In addition there are
variants of the classic outer gap model by Romani(1996)  and
Romani and Yadigaroglu (1995) who consider only the
outwardly flowing e$^{\pm}$ pairs and compute curvature, synchrotron
and inverse Compton spectra from primary and secondary particles. 

In polar cap models, primary particles are accelerated near the surface of the 
neutron star and produce $\gamma$-rays through curvature radiation and 
inverse-Compton emission.  These $\gamma$-rays convert to electron-positron
pairs in the strong magnetic field, initiating electromagnetic cascades as
the pairs radiate synchrotron $\gamma$-rays which produce further pair
generations.  The X-ray and $\gamma$-ray spectrum from such a cascade initiated
by curvature radiation was computed by Daugherty \& Harding (1996) for the
parameters of the Vela pulsar.  

As already discussed, the rotation powered pulsars emit X-rays which are a
combination of
differing amounts of three spectral components: 1)  power-law emission,
resulting from non-thermal radiation of particles
accelerated in the pulsar magnetosphere, 2) soft blackbody emission
from  surface cooling of the neutron star, and 3) a hard thermal
component from heated polar caps. In addition, spectra may
include a background of unpulsed emission
from a synchrotron nebula. Clearly, component 1) relates most directly
to the pulsar particle acceleration models. However, components 2) and
3) can also be indirectly related to the magnetospherically accelerated
particles under certain circumstances, especially if they are seen
as a pulsed component. The nebular background is due to energetic particles
injected in the surrounding nebula outside the light cylinder.
While the study of its properties is interesting in its own right
(for example, to determine the injected particle spectrum and
the strength of the nebular magnetic field
- see Harding and De Jager 1997), 
for the purposes of detecting pulsed emission, it is a background
component.

\subsection{Multi-waveband spectra and the non-thermal component of
pulsed emission}

Figure 3 shows the pulsed photon spectrum of PSR B1706-44
from EGRET detections in the
hard gamma-ray bands as well as numerous upper limits in the soft and
hard X-ray bands and TeV gamma-rays.
Since $P$ and $\dot P$ for PSR B1706-44 are very
close to those of Vela, the cascade spectrum of the two pulsars in the
Polar Cap model is expected to
be similar.  We have plotted the Vela cascade spectrum of the PC Model
(DH96 and Daugherty and Harding, in preparation) in Figure 3,
extended down to 1 keV, and normalized to the PSR B1706-44 EGRET data points.
In the Vela cascade model, the primary electrons are assumed to 
begin their acceleration above the surface
at a height of 2.0 stellar radii.
The acceleration zone extends to a height where
the electrons have achieved an
energy of 10 TeV and the cascade
begins near the top of the acceleration zone.  Acceleration 
starting at some height above the surface is likely, due to the screening
of the parallel electric field near the surface by the cascades from returning 
positrons (see Section 4.2 and Harding \& Muslimov 1998).  The cascade spectrum
is primarily synchrotron radiation from pairs, and
has a high energy turnover around 3 GeV due to pair production 
attenuation in the
local magnetic field.  
This turnover is quite sharp, because the pair production
attenuation coefficient is an exponential function of photon 
energy, and thus no pulsed
emission is expected at TeV energies.  The
spectrum has a low-energy break or turnover 
at the local cyclotron energy in the
particle rest frame, blue-shifted by the particle energy.  
In this case, the local cyclotron energy in the particles'
rest frame falls at 1.2 keV (the local field is $10^{11}$ G) and the pairs have
an average Lorentz factor of 50, giving a turnover energy of about 50 keV.
Below the turnover, the spectrum flattens into a power-law formed by 
the broadening of the
cyclotron fundamental by the pair energy distribution.  This model spectrum
falls just below the RXTE upper limit, indicating that the position of the
cyclotron energy turnover is near its minimum acceptable value.  The RXTE limit 
thus imposes a stringent upper limit on the emission radius in the polar cap
model.  For PSR B1706-44, the emission must occur within 2 stellar radii
of the neutron star surface.

In the CHRa,b model the pair production and radiation mechanisms
in the outer gap for Crab and Vela type pulsars are different
although it was shown subsequently that different angles of inclination
between the spin and magnetic axes can alter the picture substantially
(Chen and Ruderman 1993).
In the CHR model the synchrotron radiation from secondary e$^{\pm}$
pairs generated in the outer gap (by the collision of the primary
photons and the tertiary IR photons) is responsible for the 
photon spectrum in the 100 keV to 3 GeV range. The pulsed 
photon spectrum
for Vela type pulsars (with P $\ge P_t = 4.6 \times 10^{-2}
B_{12}^{2/5}$ s) in the gamma-ray region is given by Cheng and Ding
(1994) (see also Ruderman and Cheng 1988) as :

$$ \rm {d^2N_{\gamma} \over dE_{\gamma} dt} \sim 
\cases{
exp(-E_{\gamma}/E_{max}), \; \rm for \; E_{\gamma} \geq E_{max}, \cr
E^{-3/2}_{\gamma} \; \rm ln(E_{max}/E_{\gamma}), \rm \; for \; E_{min} \leq E_{\gamma}
\leq E_{max}, \cr
E^{-2/3}_{\gamma}, \; \rm for \; E_{\gamma} \leq E_{min}.
}
$$

\noindent where E$_{max}$ and E$_{min}$ are given by:

$$E_{max} \sim 3/2 \gamma_{max}^2  \rm sin \theta \; \hbar \omega_B \simeq 7.5
GeV$$
and E$_{min}$ is given by:

$$E_{min} \simeq 3/2 \gamma_{min}^2 \rm sin \theta \; \hbar \omega_B \simeq 11
MeV $$
for PSR B1706-44.
Here sin $\theta$ is the mean pitch angle for the synchrotron emitting
secondary particles and $\gamma_{max}$ corresponds to the maximum
energy of the secondary pairs, while $\gamma_{min}$ corresponds to
the secondary energy at which the radiative loss equals the
magnetospheric crossing time
and $\omega_B$ corresponds to the
local cyclotron frequency. 
The above outer gap spectrum is also plotted as a dashed line in
Fig 2 and falls well below the RXTE limit when normalized to the EGRET
points.
Specifically, Cheng and Ding quote the E$_{min}$ for 
Vela and PSR B1706-44 as (11MeV, 1.6 MeV) and E$_{max}$ as (7.5 GeV, and 0.7 TeV)
respectively. Although the E$_{min}$ of the two pulsars which have very
similar periods and magnetic fields are within an order of magnitude,
no clear explanation is given as to why the E$_{max}$ of the two pulsars
differ by nearly two orders of magnitude, especially with the same
or similar pitch angle distributions of the secondary particles.
Physically, the upper spectral break ($\hbar \omega_{max}$)
in the CHR models comes from the maximum energy of the outer gap
$\gamma$-rays  which create the energetic $e^{\pm}$ pairs whose
synchrotron radiation is in the $\gamma$-ray/X-ray regime.
The lower spectral break is related to the finite residence time
(Ruderman and Cheng 1988)
of the synchrotron radiating pairs on the open field lines
of an outer magnetosphere such that they leave the magnetosphere
(light cylinder) radius before they have a chance to radiate away
most of their energy. Since the e$^{\pm}$ leaving the light cylinder
cannot retain the memory of neutron star rotation, the radiation beyond
the light cylinder will not be pulsed, hence the 3/2 power law pulsed
synchrotron spectrum in this model is expected to be valid only
above E$_{\gamma} \geq E_{min}$ which is typically in the MeV energy
range. (CHRb obtained the spectral break between 3/2 and 2/3 power
around 100 keV, corresponding to a local magnetic field strength
of 5 $\times 10^3$ G).
Note that Ruderman and Cheng (1988) give the upper and lower
bounds as the following functions of the period of Vela-type
pulsars:
$ E_{\gamma} \leq \hbar \omega_{max} = 3 \times 10^9 (0.09s/P)^{17}
\rm eV $
and
$ E_{\gamma} \geq \hbar \omega_c \simeq 10^6 (P/0.09 s)^7 \rm eV $
respectively.

Presumably the higher value of E$_{max}$ (= 0.7 TeV)
was chosen by Cheng and Ding
(1994) 
because PSR B1706-44 does not appear to have a sharp high energy cutoff in
the EGRET range.
However, this choice of this high E$_{max}$ also violates the upper
limit of pulsed flux at somewhat lower energies around 250 GeV
due to Chadwick et al (1997). We note that, if the EGRET gamma-ray 
photon spectrum is to be explained by a 3/2 power law below 3 GeV 
due to synchrotron radiation from the secondary e$^{\pm}$
pairs in the outer magnetosphere, as in CHR or Cheng and Ding, then
the observed break in the power-law spectrum of PSR B1706-44
in EGRET data around
3 GeV might indicate an E$_{max}$ close to that of the Vela
pulsar as reported in Cheng and Ding (7.5 GeV). 

The break in the photon
spectrum at 3 GeV implies that (given a canonical maximum energy
of the primary photon of 10$^7$  MeV), the innermost
region of the magnetosphere where the production of the
``synchro-curvature" photons from the secondaries are taking place 
(see Zhang and Cheng 1997) is typically at a local magnetic field
of several times $10^6$ gauss, i.e at about 100 stellar radii (or 1/5 of the
light cylinder distance) for PSR B1706-44. For moderately inclined
pulsars, this inner boundary of the outer gap would imply an
inclination angle $\chi$ ($ cos \chi = \hat \mu . \hat \Omega$)
derived from the expression due to Halpern and Ruderman (1993):
$ r_i \simeq 4/9 cot^2 \chi (c / \Omega) $, to be around 60 degrees.

If the EGRET detected fluxes are indeed described by a 3/2 power
law below 3 GeV, then it appears that to accommodate the RXTE
upper limit obtained in this paper, one requires a break in the photon
spectrum at energies higher than about 50 keV (see Fig 3). 
As already mentioned, the PC model spectrum, if fitted to
the EGRET pulsed fluxes, cannot extend to the
keV regions unless there is a break as mentioned here.
The outer gap model's prediction due to
Cheng and Ding (1994) (which has a slope of 2/3 with a spectral
break at 1 MeV)
as shown by a dashed line in Fig 3,
passes well below the RXTE
upper limit and is not constrained by it.
 
In a modification of the outer gap model (Wang et al 1998)
the inward directed $\gamma$-ray curvature
photons (of energy $\sim$ 100 MeV, instead of the much higher energy
photons of energy $\sim 10^7$ MeV as in the CHR version) radiated by
the inward directed
primary particles produce e$^{\pm}$ pairs in the stronger
magnetic field (10$^{10}$ G) near the neutron star surface.
The radius of the magnetic pair production (at about 5 stellar radii) is
large compared to what is invoked in polar cap models. It is to be
noted that the region of pair production and subsequent X-ray and
low energy $\gamma$-ray emission is much closer to the neutron star
(and is directed starward) than the high energy emission
in the earlier CHR version of the
outer gap model. Wang et al give a photon spectral index of 3/2
with lower and upper cutoff energies at 0.1 keV and 5 MeV
respectively.
We note that such a component is unlikely to be a continuation
of the 3/2 power law photon spectrum from the EGRET energy
bands, as such an extension would violate the RXTE upper limit
as noted above and the lower cutoff of 0.1 keV can be 
ruled out for an extrapolated EGRET spectrum. 
On the other hand, such a power-law component can be present with
a lower strength, in addition to the same power law
component (due to outwardly directed particle flux)
as previously discussed by Ruderman and Cheng (1988) (see also
Cheng and Ding 1994) if the latter is taken to describe the EGRET
spectrum below $\sim 3 $ GeV. Since the former component is generated by
a starward directed beam of gamma-rays, to be observable this
X-ray component would come from the side of the
magnetosphere opposite to the observer direction.
The two components (outwardly directed X-ray and $\gamma$-ray beams
and starward directed beams) will be separated in pulsar
spin phase, possibly with different strengths in the X-ray bands
and with different spectra below $\sim$ 1 MeV.
These could be seen in phase resolved spectra should such spectra
be available in the future.

\subsection{Polar cap heating by returned particles in PC and OG
models \& X-ray spectrum}

In Fig 4 we have plotted an expanded view of the X-ray spectrum
expected from the non-thermal emission from the PC and OG models,
as discussed above, and the thermal
radiation from the heated polar cap due to returned energetic particles
from the acceleration regions. 
Also shown are the upper limits to the pulsed fluxes from ROSAT PSPC,
ROSAT HRI, ASCA, XTE and OSSE (see Thompson et al 1996 for the
references to these works; the ROSAT HRI limit is due to
Finley et al, 1998). The ROSAT PSPC upper limit corresponds
to the 18\% modulation of the total flux reported by Becker et al (1995)
(with a reported spectral index of $\alpha = 2.4$ in the 0.1 - 2.4 keV
band) while the ROSAT HRI upper limit corresponds to the upper limit of
the pulsed flux as reported in Finley et al (1998) in the same band
(we assume the same photon spectral index here as in Becker et al. (1995)).
The ASCA upper limits are due to the fluxes
in the 0.7 - 2.0 keV and 2.0 - 10.0 keV bands
and their photon index of the combined SIS and GIS
data ($\alpha$ = 1.7) as reported in Finley et al (1998). 
The RXTE upper limit in the 9.0 to 18.5 keV regime reported here is 
centered on a photon energy of 12.8 keV.
The OSSE upper limit is due to Schroeder et al (1995).

The Polar Cap and the Outer Gap models have different predictions
of the returned e$^{\pm}$ particle flux reaching the surface of the 
neutron star. If the particle fluxes indeed translate proportionally to
the {\it observable} X-ray luminosities, then
these models could potentially be distinguished
if the associated hot thermal pulsed component in the X-rays from the  
heated polar caps are detected and are distinguishable
from the the soft thermal component due to the cooling neutron
star. In the Polar Cap model,
an accurate determination of the
fraction of backflowing positrons generated in the pair formation fronts
requires a self-consistent calculation of
the screening of the electric field component parallel to the local magnetic
field due to the cascade pairs and its feedback on the pair spatial
distribution. The generated positrons which are turned around  
slightly suppress the voltage all the way down to the bottom of the
polar magnetic flux tube, so that the accelerating electric field
(${\bf E \cdot \hat B}$) component vanishes at a height 
above the
stellar surface rather than at the actual surface.
This is effectively  an enhancement of the maximum number of
electrons to be injected into the acceleration region. There is a  maximum
possible power in the backflowing positrons that will not shut off
the parallel electric field. This will lead to the  
polar cap heating luminosity estimated by Harding  and Muslimov (1998;
HM98) as:
$$ L_{e^{\pm}}^{max} \simeq 
\lambda_{max}
L_{sd} $$

\noindent where L$_{sd}$ is the spin down power of the pulsar and 
$\lambda_{max} \simeq 
(3 \times 10^{-4} - 2 \times 10^{-2})(\kappa/0.15)^2$,
and $\kappa = \epsilon I /M R^2$, I, M, R being the moment of inertia,
mass and radius of the neutron star and 
$\epsilon \sim O(1)$.
If the X-ray emission is beamed into a solid angle of
4$\pi$ steradian, the
X-ray emission due to returning particle flux gives rise to a hot polar
cap temperature of:


$$ T_{pc} = \left[{\lambda_{max} L_{sd} \over \sigma 4 \pi R^2}\right]^{1/4} $$

\noindent where R is the radius of the heated polar cap.
Taking R = $R_0 \theta_{pc} = 4.5 \times 10^4 \rm cm$, the standard
polar cap
radius for the period of PSR B1706-44, and $\lambda_{max} = 3 \times
10^{-4}$,
the backflowing positrons in the polar cap model
are expected to produce 
a luminosity of 
$1.0 \times 10^{33} \; \rm erg \; s^{-1}$  beamed into a solid angle of 4$\pi$  
steradian with a polar cap temperature of 5 $\times 10^6$ K.
In Fig 4, we have plotted the blackbody curve corresponding
to $\lambda = 3 \times 10^{-4}$ 
for the heated polar cap (hard) thermal radiation
- the maximum flux
expected from PSR B1706-44 on the basis of returned positron current as in
HM98.
This curve lies significantly above the ROSAT HRI and ASCA limits,
implying a returning particle luminosity fraction much less than the
maximum theoretically allowable.

Wang et al (1998) have argued that the X-ray emission properties
of $\gamma$-ray pulsars differ from those that are not observed
to be strong $\gamma$-ray emitters. 
For the gamma-ray pulsars,
with accelerators presumably at very many stellar radii
(``outer gap") above the neutron star surface, curvature radiation
gamma-rays moving starward produce a dense blanket of e$^{\pm}$ pairs
on closed field lines around the star.
The pair blanket affects
the X-ray emission properties by a) producing a nonthermal
synchrotron spectrum with photon power law index of 3/2 between 0.1 keV 
$\leq E_X \leq$ 5 MeV, as discussed above,
whose intensity (typically $\sim$ 10$^{33} \rm \; erg
\; s^{-1}$) depends upon the backflow current and b) forming
a cyclotron resonance blanket around the star, which shields the direct
observation of hard X-ray thermal emission from the heated polar cap
except through two narrow holes along the open field lines threading
the polar caps.
Less strongly modulated, softer X-rays can also escape by diffusing
through the cyclotron resonant blanket. The ratio of the hard
radiation flux from the heated polar cap coming through the holes (f$_h$)
to the flux through the blanket itself is given by Wang et al (1998) as:
$ f_h / f_b = r \Omega (\tau_b + 1)/2c $, where the expected optical
depth $\tau_b$ is roughly 10-200 and the blanket area is roughly $10^3$
times greater than the hole size. Therefore the radiation flux leaking
through the holes may be 10$^{-2}$ to 0.2 times the more uniformly
distributed softer flux escaping from the blanket. 
With constant bombardment of the relativistically inflowing
particles at a rate of 10$^{32} \rm s^{-1}$ (the Goldreich-Julian
current)
which radiate away much of their energy before they reach
the polar cap,
Wang et al estimated a total X-ray luminosity of
$L_x = f E_f \dot N_0 \simeq 2.1 \times 10^{32} \rm erg \; s^{-1}$, where
the factor f is the fraction of Goldreich-Julian current $\dot N_0$
that flows back through the 
outer gap accelerator to the polar cap
(taken to be 1/2 for $\gamma$-ray
pulsars not too far from their death lines). 
Here, $E_f = mc^2 \left[ 2\Omega e^2/(mc^3) \rm ln(r_{LC}/R)\right]^{-1/3}
\sim 5.5 \rm \; erg $, is the residual energy of the charged particles
impacting the polar cap.
Using this luminosity and the 
above range of the
fractional flux ratio $f_h/f_b \simeq (0.01 - 0.2)$ for
the open
field lines bundle for PSR B1706-44, the resultant hard X-ray (thermal)
component, should the beam intercept the observer, would translate
to a photon flux of

$$ F_{\nu} (T) = (0.01 - 0.2) \; {1.82 \times 10^{-2} \; E^2_{keV}
\over exp[E_{keV}/(T/2 \times 10^6)] -1 }$$ 
$$\rm photons \; cm^{-2} \;
s^{-1} \; keV^{-1} $$


\noindent Here, the heated polar cap temperature for PSR B1706-44 is
taken as the same as that
predicted for PSR0656+14, by Wang et al (1998), namely 
2$\times 10^6$ K. 

The range of the heated polar cap fluxes corresponding to the above
range of $f_h/f_b$ is shown as the shaded area in Fig 4.
Note that the available observational upper limits are also constraining
the extent of the heated polar cap emission in both PC and OG models
by limiting $\lambda$ or $f_h/f_b$. The upper range
of $f_h/f_b$ is ruled out by the ROSAT HRI limit requiring that
all but about 1 \% of the 
returned particle luminosity must be hidden at the hard X-ray and
redistributed at lower bands.
The ROSAT and ASCA limits restrict $\lambda$ ($\leq \lambda_{max}$) 
in the PC model to a value
less than $10^{-5}$, the observed limit for the
Vela pulsar. 

\subsection{Optical band luminosity predictions in OG model and R-band
observation}

Zhang and Cheng (1997) using the classic thin outer gap models of CHR
give an implied IR luminosity of the tertiary photons for Geminga.
Using the same 
line of argument, we can estimate that PSR B1706-44 should show
a total IR band luminosity of 
$$L_{IR} = n_{IR} \Delta\Omega R_L^2 cE_{IR}$$ 
$$ \sim  3 \times 10^{33} (P/0.102 s)
(\Delta\Omega / 1 sr)  
(E_{IR}/0.01 eV) \rm \;erg \; s^{-1}$$
\noindent should the
tertiary photons indeed play a crucial role in sustaining the e$^{\pm}$
discharges through IR-$\gamma$-ray photons collision (the latter in turn
produced by inverse Compton scattering of the same IR photons by highly
energetic primary electrons). 
Chakrabarty and Kaspi (1998) have found an upper limit to the R-band
magnitude for a region coincident with the pulsar of 18 mag,
corresponding to a flux of 2.4 $\times 10^{-13} \rm erg \; s^{-1} \; cm^{-2}$
at 0.7 $\mu$m, which translates to an R-band total luminosity
of 9 $\times 10^{31} \rm \; erg \; s^{-1}$ at the pulsar. If the spectrum
at the IR wavelengths is adequately described by a Rayleigh Jeans
spectrum, then the corresponding flux limit from the R-band upper limit
is well below the theoretical IR luminosity predicted from the
thin outer gap model. Inconsistencies between
the IR luminosity required by outer gap models and observed IR limits 
(IRAS) for other pulsars has been pointed out by Usov (1994),
Zhang and Cheng(1997).
The R-band upper limit corresponding to 18 mag, however is not 
sufficiently constraining compared to the scaling of the optical vs
gamma-ray luminosities expected in outer gap models (CHR)
as given by Usov (1994, see his equation (24)). Given a 
pulsed gamma-ray flux of 3$\times 10^{13}$ JyHz at 10$^{24}$ Hz 
(Thompson et al 1997), Usov predicts an optical flux of $F_{opt} \geq
4 \times 10^{-4} F_{\gamma} \simeq 1.2 \times 10^{-13} \rm \;erg \; s^{-1}
\; cm^{-2}$ for CHR outer gap models. 
However, the more stringent (possible) upper limit
of 21 mag quoted by Chakrabarty and Kaspi (1998) 
would violate the expected optical flux in the outer gap model.

\section{Conclusions:}

The main outcome of our work has been to provide an upper limit
of the pulsed X-ray flux in the 9-18 keV band which constrains
any non-thermal emission spectrum extending from EGRET gamma-ray bands
towards the soft X-ray bands of ROSAT. In particular, the cyclotron
turnover required for the Polar Cap model downwards of 50 keV, restricts
the height of polar cap acceleration to less than two stellar radii
above the NS surface.
On the other hand, if a break in the EGRET spectrum around 3 GeV
is to be consistent with the spectra at lower energies, then the 
boundary of the outer gap acceleration zone nearest to the neutron star
would be nearly 100 stellar radii away. In particular, our RXTE limit
would
tend to rule out a particular component of the outer gap model emission
(starwardly directed beam of curvature gamma-rays materializing in
relatively strong magnetic field to produce secondary e$^{\pm}$
that synchrotron radiate to produce the power law X-ray gamma-ray
spectra) requiring a lower cutoff of 0.1 keV, unless it is a weaker component
that is below the extrapolation from the EGRET spectrum to lower
energy bands. Future phase resolved spectra of this pulsar might therefore have
information to reveal with regard to the nature and strength of these
possibly different components tied to different mechanisms.

\section{Acknowledgments:}
We thank  V. Kaspi, M. Bailes, N. Wang and R.N. Manchester for providing
the radio ephemerides of PSR B1706-44 measured from ATNF. 
We thank
D.J. Thompson and Arnold Rots
for discussions and for the compilation of data on
this pulsar at other energy bands and the staff of
XTE Guest Observer Facility for
help with data analysis.
A.K.H. thanks
Joe Daugherty for assistance in producing the polar cap model cascade
spectrum.
A.R. acknowledges support through a Senior Research Associateship of
the National Research Council at NASA/Goddard and the RXTE Guest
Observer program.
%


\newpage
\vskip -1.0truein
\figureout{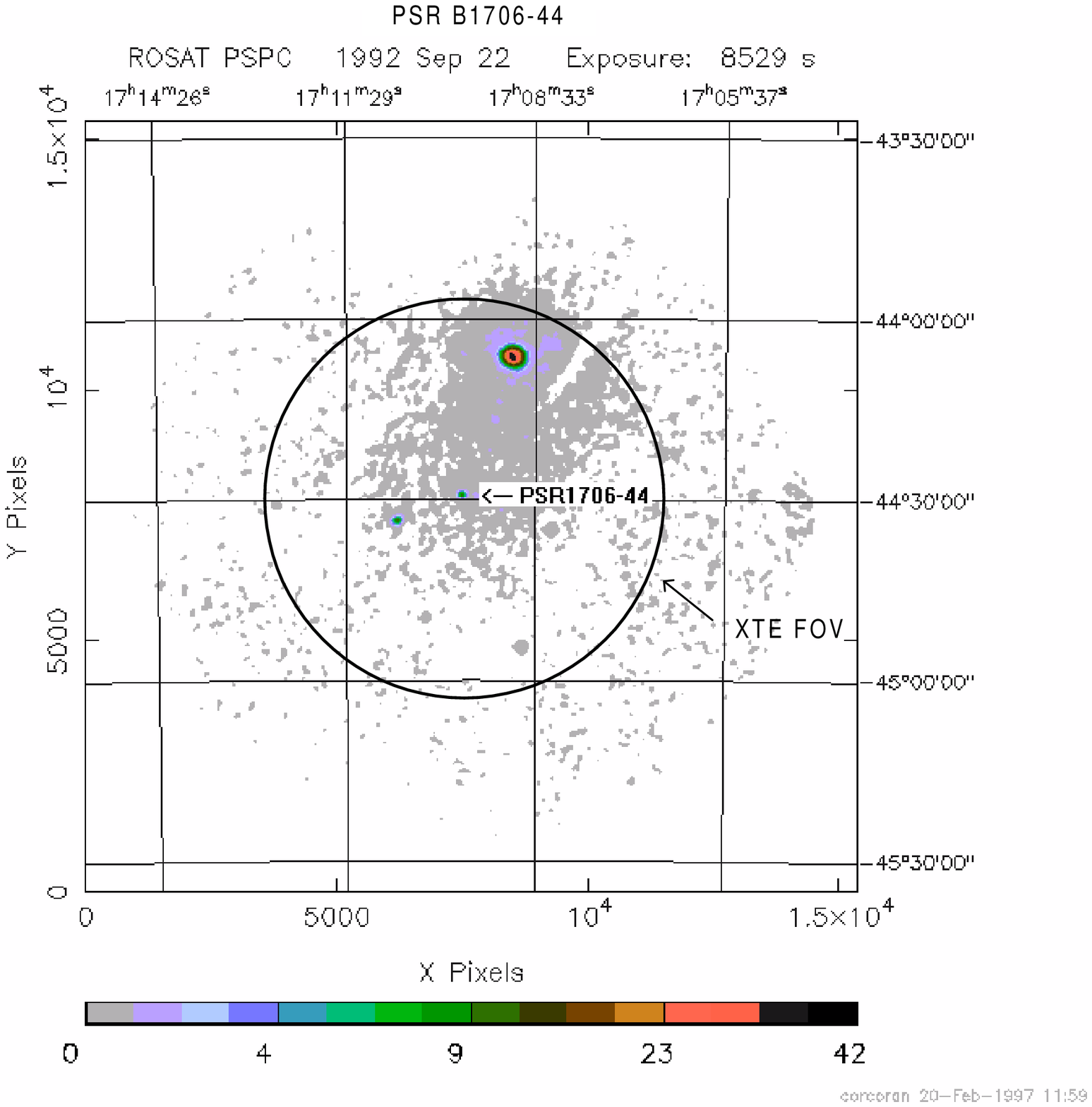}{0}{
X-ray sky around PSR1706-44 and the RXTE Field of View.
\label{F:aray:1}}
\figureout{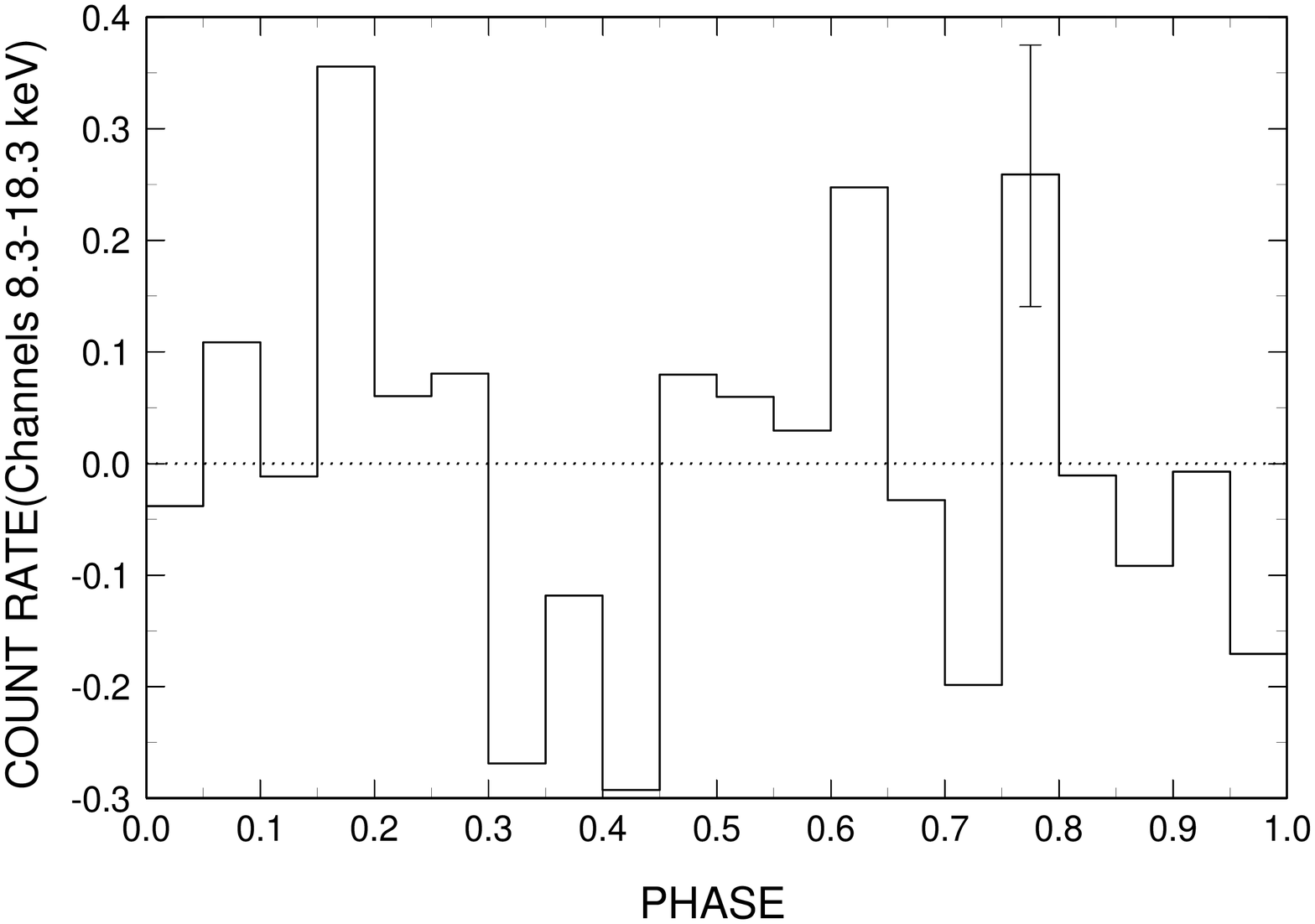}{0}{ 
PSR1706-44 phaseogram
(channels
corresponding to 8.3-18.3 keV) epoch folded according
to pulsar ephemerides. Average count rate in the phase
interval 0.0 $<$ $\phi$ $<$ 1.0 (marked as 0 on the y-axis) corresponds
to 94.2 counts/s; the 1 $\sigma$ error bar in the individual bins is
0.12 cts/s. X-ray bursts of neighboring lmxb 4U1705-44 have
been excluded from the data. Net exposure is 132ks.
\label{F:aray:2}}
\figureout{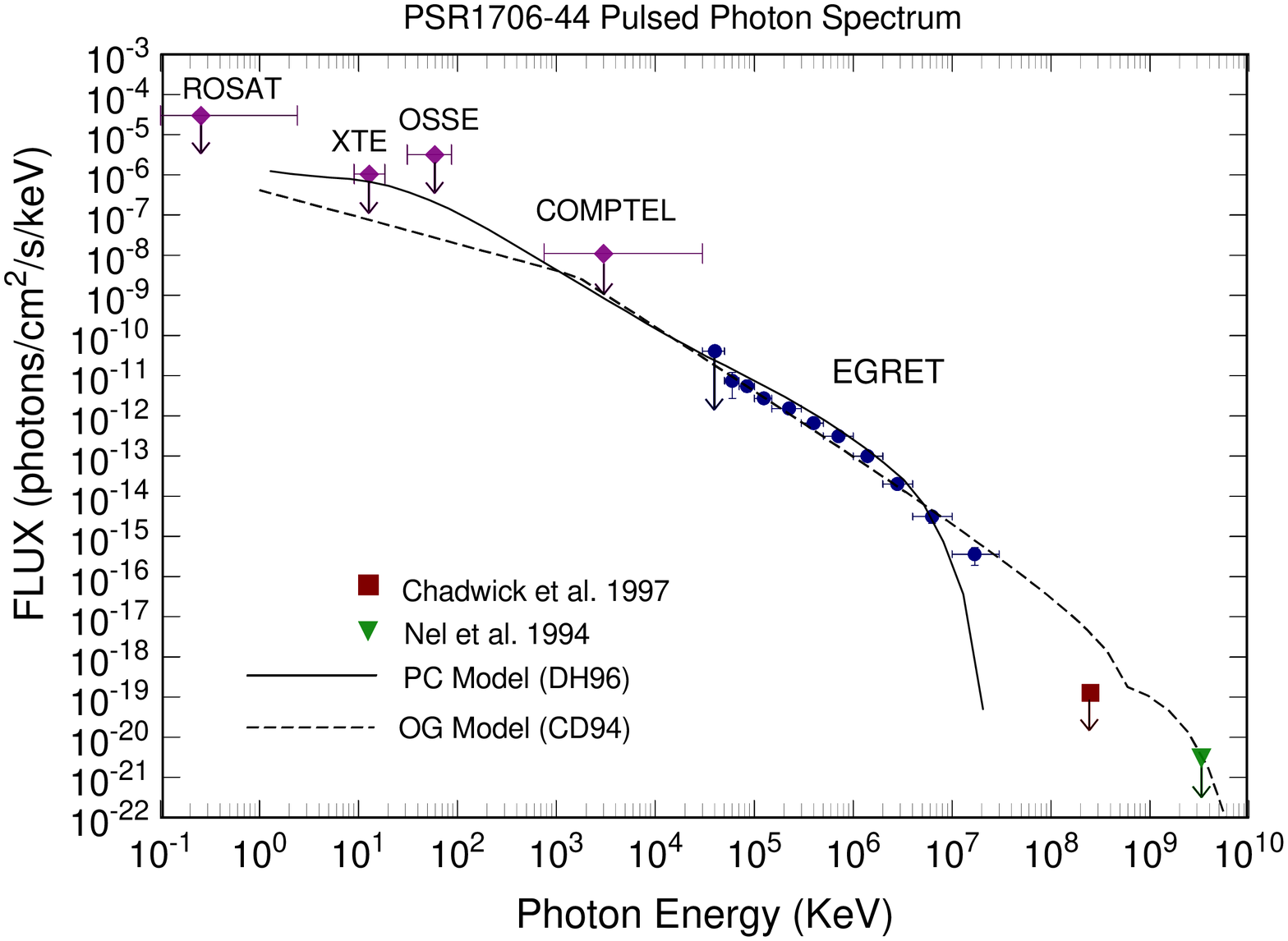}{0}{
PSR 1706-44 pulsed photon spectrum, including upper
limits, in the X-ray and gamma-ray bands. Also shown are the
predictions of the Polar Cap and Outer Gap models. For references
to the data points see Thompson et al (1996) and Finley et al (1998).
\label{F:aray:3}}
\figureout{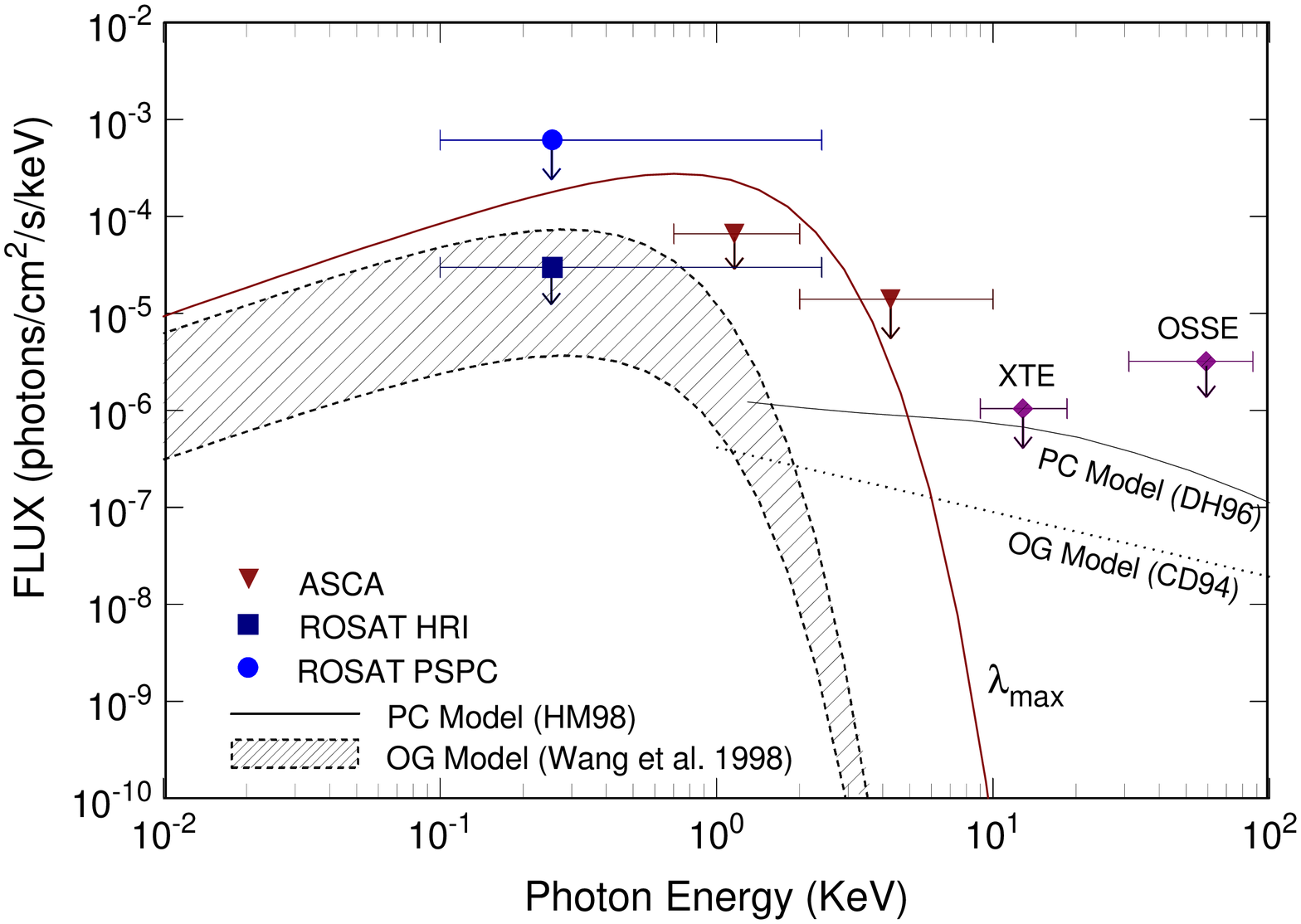}{0}{
PSR 1706-44 pulsed photon flux upper
limits, in the X-ray band. Expected blackbody spectra from
heated polar cap due to inflowing energetic particle flux
are shown for Polar Cap and Outer Gap models. Also shown are
the predictions of nonthermal emission (see Fig. 3). 
\label{F:aray:4}}
\newpage
\begin{table}
\caption{Summary of RXTE Observations of PSR B1706-44}
\label{T:aray:1}
\begin{tabular}{lccc}
   Observation Date& Beginning MJD&
   \multicolumn{1}{c}{PCA/EDS Configuration
} &
   \multicolumn{1}{c}{Total Live Time(s)}\\
\tableline
\cr
1996 Nov 9 \& 11  & 50396.371 & GoodXenon & 33487 \\

1996 Nov 10 \& 11 & 50397.922 & E\_125us\_64M\_0-1s & 52162 \\

1997 May 16 -  19 & 50584.581 & GoodXenon & 45776 \\
\end{tabular}
\end{table}

\begin{table}
  \caption{Radio Pulsar Ephemerides used in Epoch Folding$^{\dagger}$}
\begin{center}
\begin{tabular}{c|c|c|c|c} \hline
Epoch & Range of fit & $\nu_{0}$ & $\dot \nu$ & $\ddot \nu$ \\
  JD  &     JD       &   Hz      & $s^{-2}$ & $s^{-3} $   \\
\hline
50273.000000458& 50114-50433 &9.7598013942166& -8.85435D-12 & 1.47D-22
\\
50588.000000129& 50563-50613 &9.7595604761275& -8.84955D-12 & 0.00D-00
\\
\hline
\end{tabular}
\end{center}
$^{\dagger}$ Courtesy: V. Kaspi, M. Bailes, N. Wang \& R.N. Manchester;
RA,
DEC (J2000) = 17 09 42.722, -44 29 8.44 (row 1);  = 17 09 42.730,
-44 29 8.30 (row 2).
\end {table}

\end{document}